\documentclass[
superscriptaddress, twocolumn,
 amsmath,amssymb,
 aps,
 pre,
floatfix,
]{revtex4-1}

\usepackage{graphicx}
\usepackage{dcolumn}
\usepackage{braket}
\usepackage{bm}
\usepackage{amsmath}
\usepackage{physics}
\usepackage{geometry}
\usepackage{wrapfig}
\usepackage{fancyhdr}
\usepackage{mathrsfs}
\usepackage{natbib}

\begin{document}


\title{Geometry induced local thermal current from cold to hot in a classical harmonic system}

\author{Palak Dugar}
\affiliation{School of Natural Sciences, University of California, Merced, CA 95343, USA.}
\author{Chih-Chun Chien}%
\email{cchien5@ucmerced.edu}
\affiliation{School of Natural Sciences, University of California, Merced, CA 95343, USA.}


\date{\today}

\begin{abstract}
The second law of thermodynamics requires the overall thermal current to flow from hot to cold. However, it does not forbid a local thermal current from flowing from cold to hot. By coupling a harmonic system of three masses connected by a few springs to two Langevin reservoirs at different temperatures, a local atypical thermal current is found to flow from cold to hot in the steady state while the overall thermal current is still from hot to cold. The direction of the local thermal current can be tuned by the mass, spring constant, and system-reservoir coupling. The local thermal current can vanish if the parameters are tuned to proper values. We also consider nonlinear effect from the system-substrate coupling and find that the local atypical thermal current survives in the presence of the nonlinear potential. Moreover, the local atypical thermal current is robust against asymmetry of the system-reservoir coupling, inhomogeneity of the nonlinear potential, and additions of more masses and springs. In molecular or nanomechanical systems where the setup may find its realization, the direction of the local thermal current may be controlled by mechanical or electromagnetic means, which may lead to applications in information storage.
\end{abstract}

\maketitle


\section{\label{sec:level1}Introduction}
One consequence of the second law of thermodynamics is that in the macroscopic world, the overall heat flow is from a hot object to a cold one~\cite{lieb1999physics} since the entropy is non-decreasing in an isolated system. 
However, there are loopholes allowing a reversal of the direction of the thermal current from cold to hot in specific systems without violating the second law of thermodynamics. Maxwell noted the second law is "a statistical, not a mathematical, truth, for it depends on the fact that the bodies we deal with consist of millions of molecules...Hence the second law of thermodynamics is continually being violated, and that to a considerable extent, in any sufficiently small group of molecules belonging to a real body"~\cite{maxwell1878tait}. Ref.~\cite{schilling2018heat} has shown that inducing an oscillatory thermal current allows a reversal of the heat flow in the time domain, but the time-averaged behavior does not violate the second law. When a viscous electron fluid is driven by a voltage difference across a temperature gradient, it was argued that \cite{levitov2016electron} the circulation of electric current can lead to local flows from cold to hot. It was also demonstrated in \cite{micadei2017reversing} that exploiting quantum correlations between the constituents, the thermal current defined in conventional thermodynamics can flow from cold to hot without violating the second law of thermodynamics.

Classical electrodynamics requires the overall electronic current to flow from high bias to low bias~\cite{griffiths1999electrodynamics}. 
Similarly, Fourier's law  in conventional thermodynamics  describes thermal transport in macroscopic objects, formulating the thermal current to be opposite to the temperature gradient~ \cite{fourier1822theorie, dubi2011colloquium}. However, it is possible to tune the direction of particle or heat flow if no preferential direction is specified. There have been studies discussing controls of the overall direction of thermal or particle current in thermal ratchets or Brownian motors \cite{li2008ratcheting,Schwemmer2018experimental, millonas1994transport, bartussek1997ratchets, ai2004current, de2006controlled}.
However, the particle or thermal currents in those studies are not atypical because a preferred, normal direction (such as a thermal current from hot to cold in steady-state thermal transport) has not been chosen. Thus, those studies do not address whether an atypical thermal current flowing from a stationary hot reservoir to a cold one is possible.

Here we analyze a simple classical system of coupled harmonic oscillators driven by two Langevin reservoirs at different temperatures and unambiguously demonstrate a local atypical thermal current from cold to hot in the steady state. Our approach of inducing the local atypical current is based on geometric means allowing multi-paths between two points. For instance, negative differential thermal conductivity has been predicted in a triangle spin-boson system coupled to three reservoirs \cite{wang2018heat}. Ref.~\cite{lai2018tunable} shows that when the quantum electronic current flows through a multi-path triangular geometry, the current on one path may flow from low bias to high bias while the overall current always flows from high bias to low bias. 

\begin{figure}[t]
	\centering
	\includegraphics[width=0.49\textwidth]{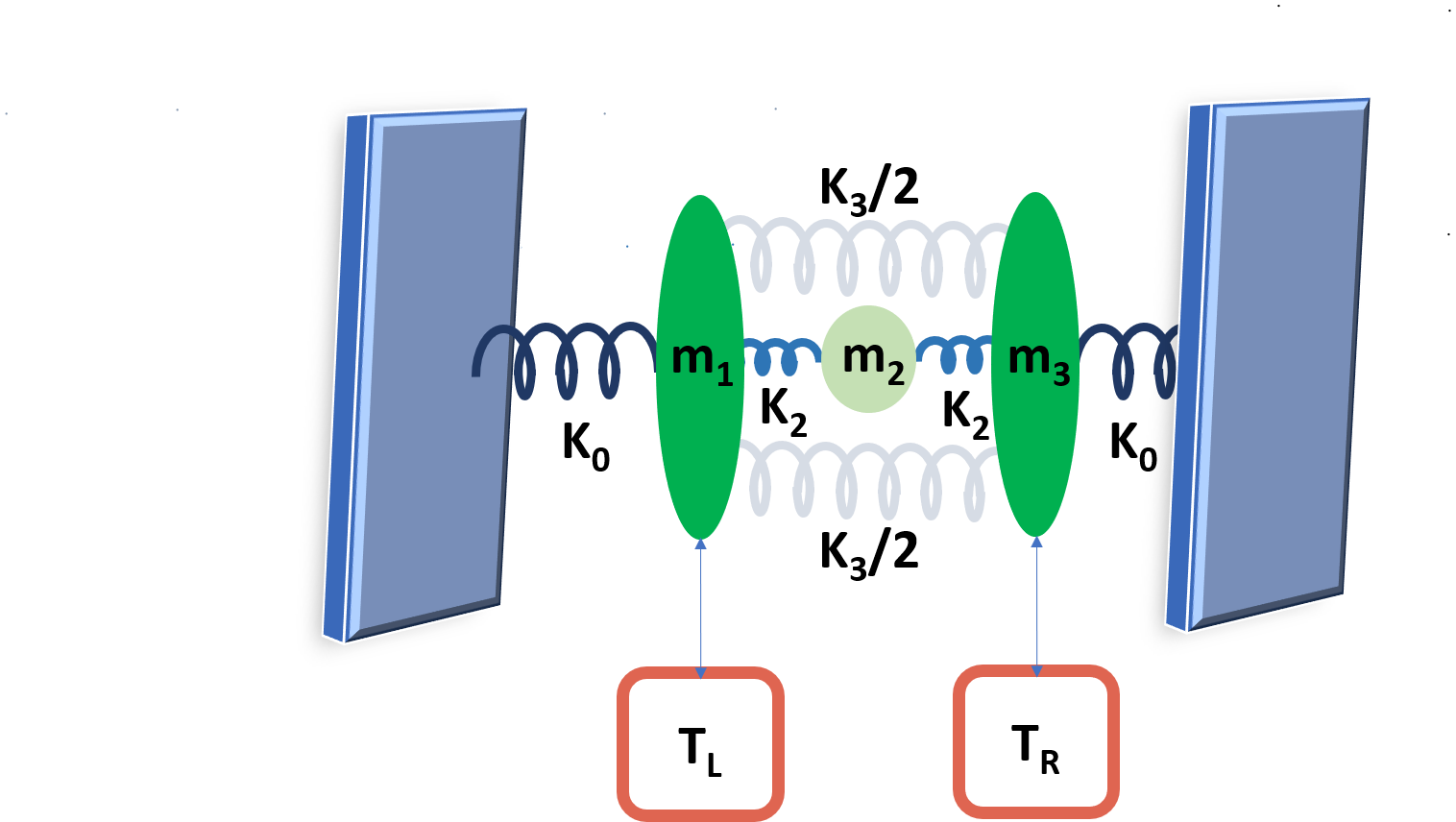}
	\caption{\label{fig:schematic} Schematic plot of the classical mass-spring system coupled to the Langevin reservoirs (red square boxes) at temperatures $T_{L}$ and $T_{R}$, respectively. $m_{1}$ and $m_{3}$ are connected to the hard walls (the slabs) via springs with spring constant $K_{0}$ and to each other with two springs with spring constant $K_{3}/2$. $m_{2}$ is connected to both $m_{1}$ and $m_{3}$ through springs with spring constant $K_{2}$.}
\end{figure}

By exploiting the analogue between quantum transport of fermions and thermal transport in classical harmonic systems \cite{kobayashi1973cp, chien2018topological}, we propose a simple multi-path geometry for thermal transport in classical harmonic systems shown in Figure~\ref{fig:schematic}. An analytic expression for the total thermal current in the steady state through the harmonic system exists~\cite{casher1971heat}, and the total current will be shown to exhibit a dip as the masses and spring constants are tuned in the regime with weak system-reservoir coupling. Following standard molecular-dynamics simulations~\cite{garcia2012noise,vanden2006second}, we will show that the dip indicates a local thermal current flows from cold to hot in the steady state after the transient behavior has decayed away. Thus, we offer a simple yet unambiguous example of Maxwell's vision~\cite{maxwell1878tait}

By analyzing the normal modes of the harmonic system and system-reservoir coupling, we will show the local atypical heat flow is a combination of the internal properties of the harmonic system and external couplings to the reservoirs. Thus, there are many knobs for tuning the local thermal current, including the mass, harmonic coupling constant, and system-reservoir coupling. Moreover, we will show that the atypical local current is robust against nonlinear onsite potentials.
One may envision molecular~\cite{chien2013tunable} or nanomechanical~\cite{chien2018topological} systems for realizing the phenomenon discussed here. 

The rest of the paper is organized as follows. In section~\ref{sec:level2} we describe the system for studying multi-path thermal transport and its modeling using the Langevin equation. Numerical methods for simulating the thermal current will be discussed. In section~\ref{sec:level3}, we present the analytic formula and its results along with the numerical results for the harmonic system driven by two Langevin reservoirs. We will show unambiguously a local atypical thermal current from cold to hot. Next, we discuss how nonlinear effects influence the local thermal currents and present phase diagrams showing where the local atypical thermal current survives. We also show the robustness nature of the atypical  of the local atypical thermal current against asymmetric system-reservoir coupling, inhomogeneous system-substrate coupling, and additions of more masses and springs. In section~\ref{sec:app}, we comment on possible ways of tuning the coupling to the reservoirs and propose applications based on the controllable local thermal current.  Section~\ref{sec:level4} concludes our work.    

\section{\label{sec:level2}Thermal transport of classical harmonic oscillators}
We consider a system of three masses $m_1$, $m_2$, and $m_3$ connected by some springs and coupled to two Langevin reservoirs at different temperatures, as illustrated in Figure~\ref{fig:schematic}. We only consider the motion of the masses in one transverse direction labeled by their displacements $x_n$ with $n=1,2,3$. The system is described by Newtonian mechanics with the Hamiltonian
\begin{equation}\label{eq:H3}
\mathscr{H}=\sum_{n} [\frac{1}{2}m_{n}\dot{x}_{n}^2 + V(x_{n}-x_{n+1}) + U(x_{n})],
\end{equation}
where $V$ is the nearest-neighbor interaction potential and $U$ is the on-site potential.
The system is confined by two hard-walls with harmonic coupling $K_0$, so one may supplement $x_0=0$ and $x_4=0$. The hard-walls prevent overall translational motions of the whole system \cite{roy2008role}. 
The results we will present only depend quantitatively on $K_0$, and all the conclusions are insensitive to $K_0$.

Here we focus on the case with only harmonic couplings between the masses, and we model the coupling to the substrate by $U(x_n)=\frac{1}{4}gx_n^4$ with coupling constant $g$ ~\cite{xiong2017crossover, savin2003heat}. Explicitly, the potentials of the three masses have the following forms: 
\begin{eqnarray}\label{eq:Potential}
(V+U)_{1}&=&\frac{K_{0}}{2}(x_{1}-x_{0})^2 + \frac{K_{2}}{2}(x_{1}-x_{2})^2 \nonumber \\ & &+\frac{K_{3}}{2}(x_{1}-x_{3})^2 + \frac{g}{4}(x_{1})^4, \nonumber \\
(V+U)_{2} &=& \frac{K_{2}}{2}(x_{2}-x_{1})^2 + \frac{K_{2}}{2}(x_{2}-x_{3})^2  \nonumber \\ 
& &+\frac{g}{4}(x_{2})^4, \nonumber \\ (V+U)_{3}&=&\frac{K_{0}}{2}(x_{3}-x_{4})^2 + \frac{K_{2}}{2}(x_{3}-x_{2})^2 \nonumber \\ & &+\frac{K_{3}}{2}(x_{3}-x_{1})^2 + \frac{g}{4}(x_{3})^4.
\end{eqnarray}
Here $K_{2}$ is the spring constant of the spring connecting $m_2$ to $m_1$ or $m_3$ and $K_{3}/2$ is the spring constant of the two identical springs connecting $m_{1}$ and $m_{3}$ directly.

Thermal transport through the system is driven by two reservoirs at temperatures $T_L$ and $T_R$. 
In this work we consider the commonly used Langevin reservoirs \cite{lepri2016heat, dhar2008heat, chen2010molecular}.
The coupled equations of motion \cite{coffey2004langevin} of the driven system are
\begin{eqnarray}\label{eq:EOM}
m_{1,3}\ddot{x}_{1,3}&=&F_{1,3} - b_{L,R}\dot{x}_{1,3} + \eta_{L,R}(t), \nonumber \\
m_2\ddot{x}_{2}&=&F_{2}.
\end{eqnarray} 
Here $F_{n}=-\partial (U+V)_{n}/\partial x_n$ is the deterministic force on the $n$-th mass.
The subscripts $L$ and $R$ denote the left and right reservoirs, respectively.  $b_{L,R}$ are the friction coefficients coupling the system to the reservoirs, and $\eta_{L,R}(t)$ are the stochastic forces from the reservoirs. $\eta_{L,R}(t)$ satisfy $\langle\eta_{L,R}(t)\rangle=0$ and 
\begin{equation}
\langle{\eta_{L,R}(t_{1})\eta_{L,R}(t_{2})}\rangle= 2bk_{B}T_{L,R}\delta(t_{1}-t_{2}),
\end{equation} 
where $k_B$ is the Boltzmann constant.
The latter relation guarantees the fluctuation-dissipation theorem \cite{kubo1966fluctuation, garcia2012noise, dhar2008heat}. In the following, we consider $m_1=m_3=m$ and $K_0=K_3=K$. Moreover, we will show that slightly different values of  $b_L$ and $b_R$ do not change the results qualitatively, so we will use the symmetric system-reservoir coupling, $b_L=b_R=b$, unless stated specifically.

The local thermal current from site $i$ to site $j$ can be defined with the help of the continuity equation~\cite{lepri2003thermal}. The general form is $J_{ij}=\langle F_{ij}\dot{x}_{j}\rangle$, where $F_{ij}$ is the force acting on mass $j$ due to mass $i$. 
For a harmonic system in the steady state, $J_{ij}=-J_{ji}$. The explicit form of the total thermal current from $m_1$ to $m_3$ is 
\begin{equation}\label{eq:J13}
\langle{{J}_{13}}\rangle=\langle{\dot{x}_{3}(K_{2}x_{2} + K_{3}x_{1})}\rangle.
\end{equation}
In the steady state, it is equal in magnitude but in the opposite direction of the thermal current from $m_3$ to $m_1$.
Similarly, we can find the local current flowing from $m_1$ to $m_2$, which is the same as the local current through $m_2$ to $m_3$ in the steady state. Explicitly, 
\begin{equation}
\langle{{J}_{12}}\rangle=\langle{\dot{x}_{2}(K_{2}x_{1})}\rangle,
\end{equation}
which is also equal to $\langle{{J}_{23}}\rangle$ in the steady state. We will numerically evaluate the local currents through each path for the harmonic system as well as the case including nonlinear effects via the onsite potential $U(x_j)$ modeling the substrate effect.

Dimensionless quantities can be constructed by using $m$, $K$, $k_{B}$, and $\hbar$. For instance, the units of energy, temperature, time, angular frequency, length, thermal current, system-reservoir coupling, and system-substrate coupling are $E_{0}=\hbar \sqrt{\frac{K}{m}}$,  $T_{0}=\frac{\hbar}{k_{B}} \sqrt{\frac{K}{m}}$,  $t_{0}=\sqrt{\frac{m}{K}}$,  $\omega_{0}=\sqrt{\frac{K}{m}}$,  $l_{0}=\sqrt{\frac{\hbar}{\sqrt{mk}}}$, $J_{0}=\hbar\frac{K}{m}$,  $b_{0}=\sqrt{mK}$, and  $g_{0}=\frac{\sqrt{mK^3}}{\hbar}$, respectively. 
Take a nano system \cite{li2004vibrational, du2018terahertz} for example, $m \simeq 10^{-26}$ kg, $\sqrt{\frac{K}{m}}\simeq 10^{12}$ rad/s, $l_{0}\simeq10^{-10}$ m, and $T_{0}$ is around $73K$. The setup of Fig.~\ref{fig:schematic} is generic and may be applicable to molecular or nano-mechanical systems in the classical regime \cite{chien2013tunable,akyildiz2010internet}, or even macroscopic objects as long as the Langevin equation~\eqref{eq:EOM} applies.

Available computer simulation methods for the Langevin equation include the first-order Euler-Maruyama method \cite{higham2001algorithmic}, the hybrid Heun method \cite{garcia2012noise}, and the complete second-order method \cite{vanden2006second}, etc. The white noise is simulated by a Weiner process. In absence of nonlinearity, the results from those methods are practically indistinguishable if the time increment is carefully chosen. To handle nonlinearity, however, the Heun or complete second-order method allows a larger step size in the simulation and gives more reliable results. 

We follow the algorithms from Refs. \cite{garcia2012noise,vanden2006second} and present the results accurate up to the second-order of the time increment.
The time step size in our simulations is  $\Delta t/t_{0}=10^{-4}$. We present the results averaged over an ensemble of 1600 independent  realizations. To ensure that the system is in the steady-state regime, we monitor the time evolution of the thermal current and wait until the transient behavior decays away. In general, we start taking the steady-state value after $t>300t_{0}$ and then average the value over a time period of $ \tau/t_{0}= 500$ afterwards. Importantly, we have checked there is no energy accumulation in the system in the steady state by verifying the thermal current coming into each mass equals the current out of each mass. For instance, $\langle J_{12}\rangle=\langle J_{23}\rangle$, $\langle J_{13}\rangle = -\langle J_{31}\rangle$, etc. Unless specified otherwise, the reservoirs were maintained at $T_{L}/T_{0}=2$ and $T_{R}/T_{0}=1$. To map out the flow direction of the thermal current through $m_2$, we vary $K_{2}/K_{3}$, $m_2/m_3$, $b_{L,R}$, and $g$ and check the steady-state current in each case.

\section{\label{sec:level3}Atypical local thermal current from cold to hot}

\subsection{Analytic formula for total current}
We will start with the case with the harmonic case with $g=0$ and investigate the thermal currents in the linear system. A general formalism for the thermal conductance of harmonic systems coupled to Langevin reservoirs can be found in Refs.~\cite{casher1971heat, dhar2008heat}. For a harmonic system consisting of $N$ masses connected by springs and coupled to two Langevin reservoirs with a temperature difference $\Delta T=T_L-T_R$ at the ends, the total thermal current is given by
\begin{eqnarray}\label{eq:CL_J}
J&=&\Delta T b_L b_R \int_{-\infty}^{\infty}\frac{d\omega}{\pi} \omega^2 |C_{lN}|^2[(K_{1,N} \nonumber \\
& & - \omega^2 b_L b_R K_{2,N-1})^{2}+\omega^2 (b_R K_{1,N-1} \nonumber \\
& &+b_L K_{2,N})^{2}]^{-1}.
\end{eqnarray}
Here one constructs the force matrix $\Phi$ specifying the harmonic couplings between pairs of masses, the mass matrix $M$ is diagonal with the masses $m_1,m_2,\cdots,m_N$ along the diagonal, and the matrix $Z=\Phi-M \ddot{\omega}-i\omega B$, where $B$ has only two nonvanishing elements $B_{11}=b_L$ and $B_{NN}=b_R$. Then, $K_{i,j}$ denotes the determinant of the matrix from the $i$-th row (column) to the $j$-th row (column) of the matrix $(\Phi-M\omega^2)$. $C_{1N}$ is the cofactor of the $(1,N)$-th element of $Z$. For the setup shown in Fig.~\ref{fig:schematic}, the force matrix is
\begin{equation}
\Phi=\left(\begin{array}{ccc} 
K_0+K_2+K_3 & -K_2 &-K_3 \\
-K_2 & 2K_2 &-K_2 \\
-K_3& -K_2 & K_0+K_2+K_3
\end{array}
\right).
\end{equation}
The numerical value of the total thermal current $J_{13}$ flowing through the device shown in Fig.~\ref{fig:schematic} can be obtain from Eq.~\eqref{eq:CL_J} after one obtains the cofactor and determinants needed for the evaluation.

\begin{figure}[th]
\includegraphics[width=\columnwidth]{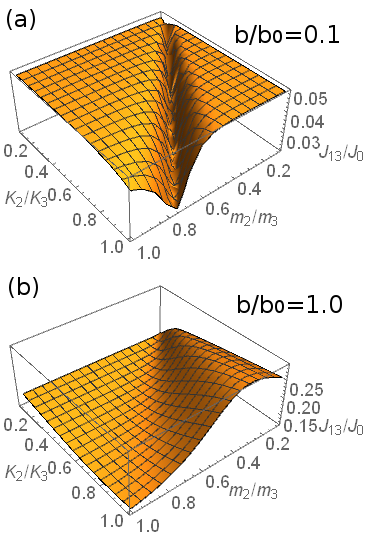}
\caption{Total thermal current $J_{13}$ through the harmonic system shown in Fig.~\ref{fig:schematic}, according to the analytic formula, Eq.~\eqref{eq:CL_J}. Here $b_L=b_R=b$, $m_1/m_3=1$, $K_0/K_3=K_1/K_3=1$, $g=0$, $\Delta T/T_0=1$, and $b/b_0=0.1$ for (a) and $b/b_0=1.0$ for (b). }
\label{fig:analytic}
\end{figure}

Figure~\ref{fig:analytic} shows the total thermal current $J_{13}$ as a function of $m_2/m_3$ and $K_2/K_3$ according to Eq.~\eqref{eq:CL_J} with $m_1=m_3$, $K_0=K_1=K_3$, $\Delta T/T_0=1$, and selected values of $b_L=b_R=b$. When $b$ is small, for instance $b/b_0=0.1$, one observe the surface of $J_{13}$ exhibits a dip, implying a non-monotonic dependence of $J_{13}$ on the parameters $m_2/m_3$ and $K_2/K_3$. However, the dip disappears when $b$ increases. We have checked the more general cases with $0.5 \le b_L/b_R \le 2$, and the total current behaves qualitatively the same. Moreover, the observation of a dip (or no dip) when $b_L,b_R$ are small (or large) remains. At this stage, the dip of $J_{13}$ may look mysterious. In the next section, we will show that, by numerically analyzing the local thermal current through each path, the dip of the total thermal current is associated with a local atypical thermal current flowing through the mass $m_2$ from cold to hot.

Incidentally, simplified expressions based on Eq.~\eqref{eq:CL_J} are available for 1D  harmonic chains with certain patterns in the infinite-chain limit~\cite{casher1971heat,Pereira11,Kannan12,chien2017thermal}. The analytic formula also helps clarify the influence of topological edge states on thermal transport in 1D harmonic systems~\cite{chien2018topological}.

\begin{figure}[t]
\includegraphics[width=0.9\columnwidth]{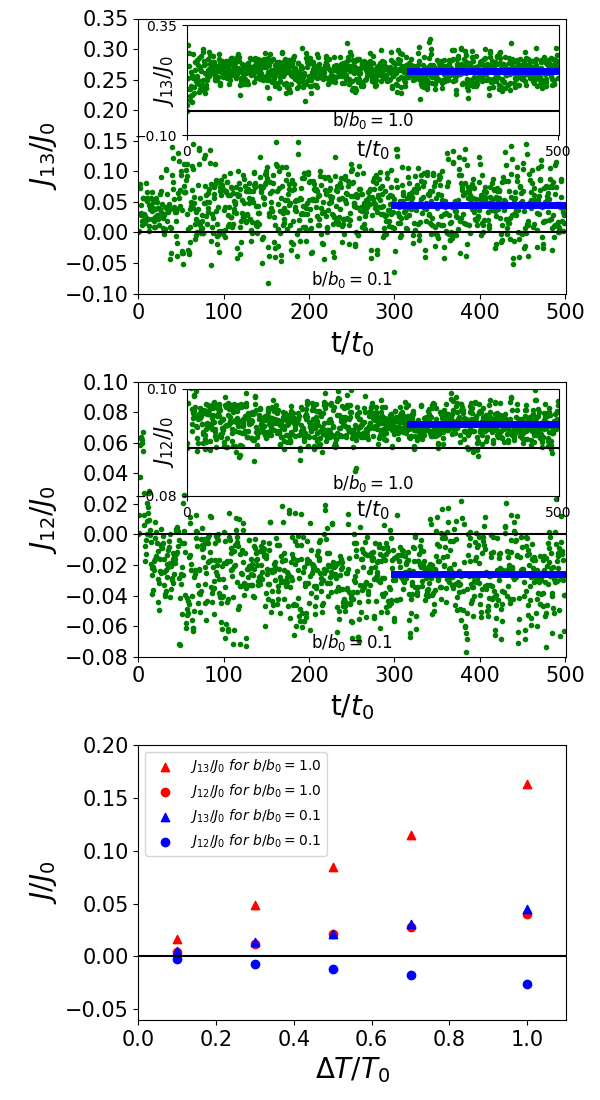}
	\caption{\label{fig:SScurrent} The total thermal current $J_{13}$ (top panel) and the local thermal current $J_{12}$ through $m_2$ (middle panel) for $b/b_{0}=0.1$, showing the local atypical current from cold to hot. The insets present the corresponding quantities for $b/b_{0}=1.0$, showing all currents flowing from hot to cold. The dots show the average over $1600$ realizations. The thick blue lines show the average over a period of $500 t_{0}$ in the steady state ($t>300t_0$). Here $K_2/K_3=0.35$, $m_2/m_3=0.3$, $\Delta T/T_0=1$, and $g/g_{0}=0$. 
	The bottom panel shows the total and local thermal currents vs. $\Delta T$ for the two cases shown above.}
\end{figure}

\subsection{\label{sec:level3a}Atypical local current in harmonic System}
The anomalous behavior of the total thermal current in Fig.~\ref{fig:analytic} actually encompasses an interesting phenomenon of a local atypical thermal current flowing opposite to the total current. We start with the linear system by setting the nonlinear substrate coupling $g$ to zero and use numerical simulations to obtain the local thermal current. Figure~\ref{fig:SScurrent} shows the atypical behavior of a selected example with $b_L=b_R=b=0.1 b_{0}$. 

The top and middle panels of Fig.~\ref{fig:SScurrent} show the total current $J_{13}$ from the hot reservoir to the cold one and the local current $J_{12}$ through $m_2$, respectively. The opposite directions of $J_{13}$ and $J_{12}$ in the steady state for the case with $b/b_{0}=0.1$ demonstrate unambiguously the existence of a local atypical thermal current from cold to hot. We emphasize the steady-state values are taken after $t>300t_0$ to ensure the transient behavior has decayed away. In the inset, we show the typical, normal behavior of another example with $b/b_{0}=1.0$, where $J_{12}$ and $J_{13}$ show the same sign in the steady state. We found that, in general, the local thermal current through $m_2$ can flow from cold to hot only in the weakly coupling regime when $b/b_{0}$ is small compared to the other parameters.
The dependence of the thermal current of a one-dimensional (1D) harmonic system on $b$ has been studied in Refs.~\cite{lepri2003thermal,velizhanin2015crossover}, but the setup shown in Fig.~\ref{fig:schematic} is not a simple 1D system.

To verify the local thermal current from cold to hot is not an artifact, we vary the temperature difference $\Delta T$ between the two reservoirs and show in the bottom panel of  Figure~\ref{fig:SScurrent} the local current from cold to hot indeed scales linearly with $\Delta T/T_{0}$ just like the total current. 
The result thus establishes the existence of a local atypical thermal current from cold to hot in a simple classical harmonic system driven by Langevin reservoirs.
By sweeping the values of $(K_{2}/K_{3}, m_{2}/m_{3})$ in the parameter space when $b/b_{0}$ is small, we found the steady-state thermal current through $m_2$ can be either in the direction from $m_1$ to $m_3$ (from hot to cold) or vice versa (from cold to hot) when $T_L$ and $T_R$ are fixed.

\begin{figure}[th]
	\centering
	\includegraphics[width=0.49\textwidth]{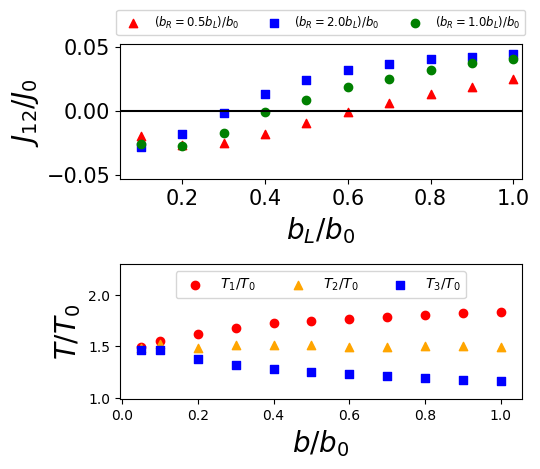}
	\caption{\label{fig:HarmonicPhase} (Top panel) The dependence of the local current $J_{12}$ on the system-reservoir coupling $b_L/b_0$ for asymmetric couplings $b_R/b_0=0.5b_L/b_0$ (triangles) and $2.0b_L/b_0$ (squares) and symmetric coupling $b_R/b_0=b_L/b_0$ (circles).
	(Bottom panel) The local temperatures of the three sites as functions of $b/b_0$ for the case of symmetric coupling in the steady state. Here $m_{2}/m_{3}=0.3$, $K_{2}/K_{3}=0.35$, $\Delta T/T_0=1$, and $g=0$.}
\end{figure}

Away from the small $b/b_{0}$ regime, the system only exhibits the normal behavior for reasonable values of $m_2/m_3$ and $K_2/K_3$. In the insets of Fig.~\ref{fig:SScurrent}, we present the total and local thermal currents of the case with the same parameters except $b/b_{0}=1.0$, showing all the currents flow from hot to cold. 
The dependence of $J_{12}$ on $b$ is illustrated in the upper panel of Fig.~\ref{fig:HarmonicPhase} with the selected parameters $m_{2}/m_{3}=0.3$ and $K_{2}/K_{3}=0.35$. For symmetric system reservoir coupling, $b_R/b_0=b_L/b_0$ as $b$ increases, $J_{12}$ changes from the atypical (cold to hot) direction to the normal (hot to cold) direction. Importantly, there is a critical point ($b/b_{0}\approx 0.4$ for this case), where the local current $J_{12}$ vanishes in the steady state. We note that there is still an overall thermal current flowing through the two springs coupling $m_1$ and $m_3$ directly, but the path through $m_2$ carries no thermal current in the steady state.
 
Asymmetric system-reservoir couplings with $b_R/b_L$ of order $1$ yields qualitatively the same results as the case with symmetric system-reservoir coupling. As shown in the upper panel of Fig.~\ref{fig:HarmonicPhase}, the local current $J_{12}$ can change from the atypical to normal behavior as $b_L$ increases if we choose $b_R=0.5b_L$ or $b_R=2b_L$.

We also evaluate the local temperatures of the three masses and verify that $m_1$ is really hotter than $m_2$ in the steady state. The local temperature of mass $j$ (with $j=1,2,3$) is defined as
\begin{equation}
T_j=\frac{1}{k_B}m_j \langle v_j^2 \rangle.
\end{equation}
In the lower panel of Fig.~\ref{fig:HarmonicPhase}, we show the steady-state local temperatures of the three masses as a function of $b$ for the case with symmetric system-reservoir coupling. One can see that regardless of the presence the atypical local thermal current, the local temperatures always follow $T_1 >T_{2}>T_3$. The local temperatures thus firmly establish the atypical behavior of the local thermal current because it can indeed flow from cold to hot if the system and system-reservoir coupling are tuned to the right parameters.

\begin{figure}[th]
	\centering
	\includegraphics[width=0.49\textwidth]{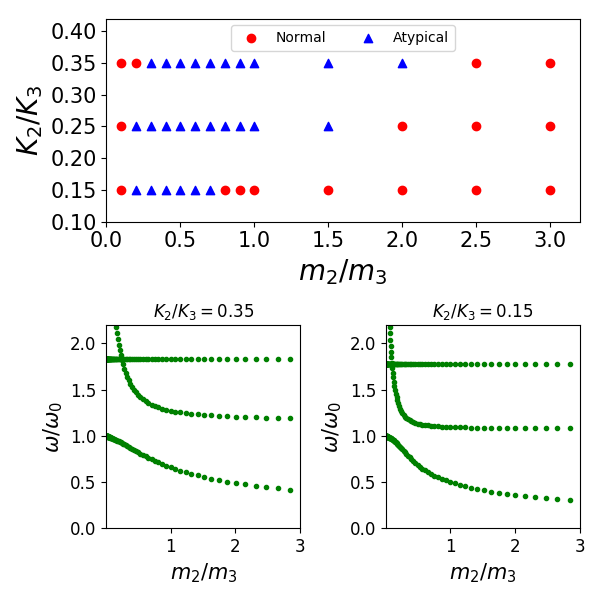}
	\caption{\label{fig:Eplot}  (Top panel) Phase diagram showing where a local atypical thermal current from cold to hot can be found. The blue triangles (red dots) indicate where the local current is atypical (normal). Here $b/b_{0}=0.1$ and $g/g_{0}=0$. The bottom panels show the normal mode frequencies, $\omega/\omega_{0}$, of the system shown in Fig.~\ref{fig:schematic}  without the reservoirs for  $K_2/K_3=0.35$ (left)  and $K_2/K_3=0.15$ (right).}
\end{figure}

The upper panel of Figure~\ref{fig:Eplot} shows the phase diagram of the system with $b/b_0=0.1$, where the blue triangles $($red dots$)$ indicate where a local atypical $($normal$)$ thermal current through $m_2$ is observed when $b/b_0=0.1$. Increasing $b/b_0$ in the atypical regime always drives the system from one with a local atypical current to one with only normal currents, similar to the result shown in the upper panel of Fig.~\ref{fig:HarmonicPhase}.
On the other hand, varying $m_2/m_3$ and $K_2/K_3$ leads to more complicated behavior of the local current and the atypical regime may be sandwiched in between the normal regimes.

The dependence of the direction of the local thermal current on the system-reservoir coupling $b$, shown in Fig.~\ref{fig:HarmonicPhase}, indicates the mechanism behind the atypical local current is not an intrinsic property of the harmonic system.
To corroborate the observation, we evaluate the normal-mode frequencies of the harmonic system shown in Fig.~\ref{fig:schematic}  without the reservoirs and show the spectra in the bottom panels of Fig.~\ref{fig:Eplot} for two selected cases.

As shown in the bottom panels of Figure~\ref{fig:Eplot}, the normal-mode frequencies for both values of $K_2/K_3$ exhibit level crossings. We found the locations of the level crossings are close to the left boundary of the region exhibiting the atypical current in the phase diagram shown in the top panel of Figure~\ref{fig:Eplot}. On the other hand, no additional level crossing appears as $m_2/m_3$ increases, but the phase diagram shows that the system changes from the atypical regime to the normal regime when $m_2/m_3$ becomes large. 
With the observation that tuning the system-reservoir coupling $b$ also controls the direction of the local thermal current even when $m_2/m_3$ and $K_2/K_3$ are fixed, as shown in Fig.~\ref{fig:HarmonicPhase}, the atypical current cannot be solely attributed to the normal-mode spectrum of the harmonic system. Instead, it is a combined effect of the system and reservoirs. 

Moreover, the location of the dip in the total current from the analytic formula, shown in Fig.~\ref{fig:analytic}, is close to the left boundary of the atypical regime shown in Fig.~\ref{fig:HarmonicPhase} for small $b/b_0$. We found this to be generic and one can use the dip in the total current from the analytic formula to estimate where the atypical local current emerges. There is no indication of the right boundary of the atypical regime from the analytic formula, though. Therefore, it is insufficient to determine the whole atypical regime by analyzing the total thermal current or the normal-mode spectrum. One has to numerically analyze each local current as we did here.

For quantum transport of electrons through a triangular quantum dot metastructure~\cite{lai2018tunable}, the local atypical electric current is due to the wave nature of quantum particles. Thermal transport in classical harmonic systems is carried by the normal modes coupled to the reservoirs. The normal modes may be viewed as mechanical waves. As the reservoirs pump in and take out energy through the normal modes, there is no rule forbidding a path from overshooting the overall thermal current. Moreover, the conservation of charge in electronic transport imposes Kirchhoff's law requiring the net electric current through a node should be zero~\cite{griffiths1999electrodynamics}. Similarly, in steady-state thermal transport the net thermal current through a mass in a harmonic system should vanish. Therefore, if an overshoot occurs along a path, another path will compensate for the excess thermal current by carrying the thermal current backward, resulting in the atypical local thermal current.

\subsection{\label{sec:level3b}Atypical local current and nonlinear onsite potential}
The harmonic system with the Langevin reservoirs is a linear system, allowing a detailed analysis of its dynamics. After establishing the local thermal current from cold to hot, we consider an effective nonlinear onsite potential $U(x_n)=(1/4)gx_n^4$ modeling the coupling between the masses and the substrate following Refs.~\cite{xiong2017crossover, savin2003heat}. In the presence of the nonlinear potential, the vibrational spectrum of the system cannot be described by the normal modes. Nevertheless, our simulations of the system with the nonlinear substrate effect using the second order method can test the robustness of the  local thermal current from cold to hot against the onsite nonlinear potential.

\begin{figure}[th]
\centering
\includegraphics[width=0.49\textwidth]{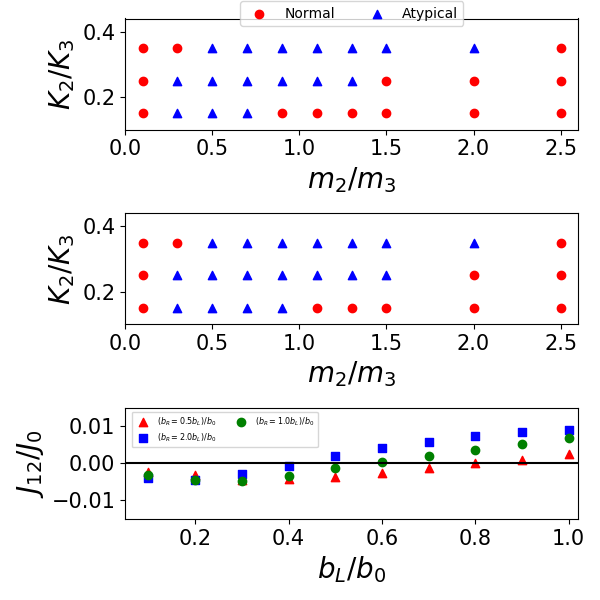}
\caption{\label{fig:NonlinearDiagram} Top and middle panels: Phase diagrams of the system with nonlinear substrate effect. Here $g/g_{0}=0.1$ for the top panel and $g/g_{0}=0.5$ for the middle panel. The blue triangles (red dots) show where a local thermal current from cold to hot can (cannot) be observed. Here $b/b_{0}=0.1$. Bottom panel: The local thermal current as a function of the system-reservoir coupling for the symmetric case  with $b_L=b_R$ (circles) and asymmetric cases with $b_L=0.5b_R$ (triangles) and $b_L=2.0b_R$ (squares).  Here $m_2/m_3=0.9$, $K_2/K_3=0.35$, $\Delta T/T_0=1$, and $g/g_0=0.5$.
}
\end{figure}

The top and middle panels of Figure~\ref{fig:NonlinearDiagram} show the phase diagrams of the system with $g/g_{0}=0.1$ and $0.5$ at fixed $b/b_{0}=0.1$. As $g/g_0$ increases, the regime with a local atypical thermal current (the blue triangles) remains almost the same. Therefore, the atypical local thermal current is robust against the nonlinear onsite potential.
On the other hand, the dependence on $m_2/m_3$ and $K_2/K_3$ shows similar behavior as the case with $g/g_0=0$. 
The bottom panel of Figure ~\ref{fig:NonlinearDiagram} shows the dependence of the local thermal current $J_{12}$ on the system-reservoir coupling for the symmetric ($b_L=b_R$) and asymmetric ($b_L\neq b_R$) cases for $m_2/m_3=0.9$, $K_2/K_3=0.35$, $\Delta T/T_0=1$, and $g/g_0=0.5$. As one can see, the local thermal current can be tuned from the atypical regime to the normal regime by increasing the system-reservoir coupling in both the symmetric and asymmetric cases.

\subsection{Robustness of atypical local current}
We have shown that the atypical local thermal current in the steady state is robust against asymmetric system-reservoir coupling ($b_L\neq b_R$) in both the linear ($g/g_0=0$) and non-linear ($g/g_0=0.5$) cases, as demonstrated in the top panel of Fig.~\ref{fig:HarmonicPhase} and the bottom panel of Fig.~\ref{fig:NonlinearDiagram}. 
Next, we found the atypical thermal current survives after introducing inhomogeneity to the nonlinear system-substrate coupling. Explicitly, we consider $U(x_n)=(1/4)g_n x_n^4$ with tunable $g_n$ for each mass. For the extreme case with $g_1=g_3=0$ and $g_2/g_0=0.5$, the phase diagram is still similar to the one with $g_n/g_0=0.5$ for all $n$ shown in the lower panel of Fig.~\ref{fig:NonlinearDiagram} with the same $b/b_0$. The robustness of the atypical thermal current against inhomogeneous nonlinear onsite potential is in contrast to the atypical local electronic current in the triangular quantum-dot metastructure of Ref.~\cite{lai2018tunable}. The latter was shown to be sensitive to inhomogeneous onsite interactions.

\begin{figure}[th]
\centering
\includegraphics[width=0.49\textwidth]{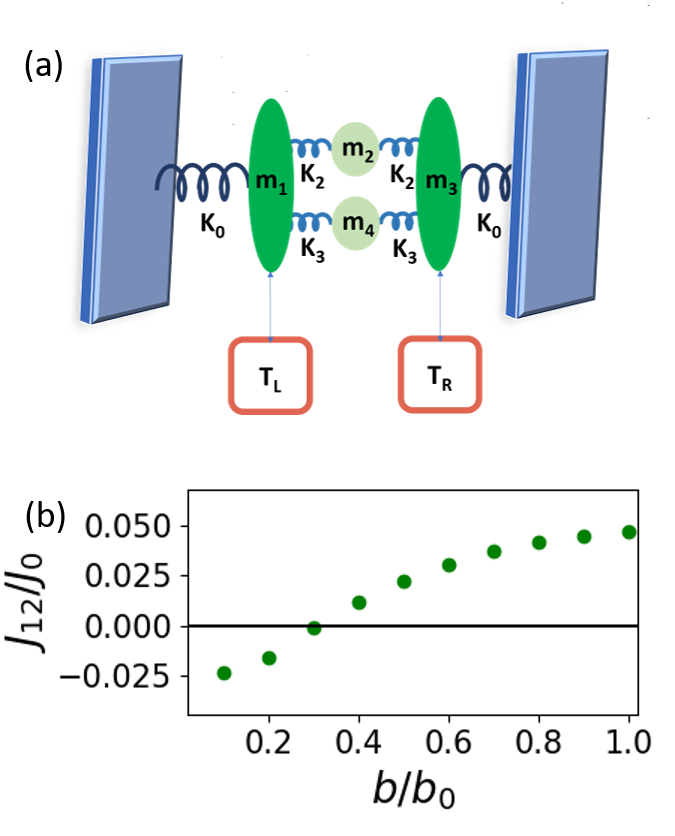}
\caption{\label{fig:foursite} (a) Illustration of a 4-mass setup. The addition of $m_4$ avoids a direct coupling between $m_1$ and $m_3$, which are connected to the Langevin reservoirs at temperatures $T_{L}$ and $T_{R}$, respectively.  The convention follows Fig.~\ref{fig:schematic}.
(b) The local thermal current $J_{12}$ through $m_2$ in the steady state as a function of the symmetric system-reservoir coupling  $b_R=b_L=b$. Here $m_1/m_3=1$, $m_2/m_3=0.4$, $m_4/m_3=0.5$, $K_0/K_3=1$, $K_2/K_3=0.35$, $\Delta T/T_{0}=1$, and $g=0$.
}
\end{figure}

Another test of the atypical local current is to introduce a more complicated setup, where the direct coupling between $m_1$ and $m_3$ in Fig.~\ref{fig:schematic} is avoided. The setup is illustrated in Fig.~\ref{fig:foursite} $(a)$, where two paths with $m_2$ and $m_4$ in the middle, respectively, connecting $m_1$ and $m_3$. The Hamiltonian is given by
\begin{equation}
H_4=\sum_{n=1}^{4} [\frac{1}{2}m_{n}\dot{x}_{n}^2 + V(x_{n},x_{n+r}) + U(x_{n})],
\end{equation}
where $V$ and $U$ denote the harmonic coupling and onsite potentials, similar to Eq.~\eqref{eq:H3}. Two hard-walls are introduced with $x_0=x_5=0$. The potentials have the following forms: 
\begin{eqnarray}\label{eq:4Potential}
(V+U)_{1}&=&\frac{K_{0}}{2}(x_{1}-x_{0})^2 + \frac{K_{2}}{2}(x_{1}-x_{2})^2 \nonumber \\ & &+\frac{K_{3}}{2}(x_{1}-x_{4})^2 + \frac{g}{4}(x_{1})^4, \nonumber \\
(V+U)_{2} &=& \frac{K_{2}}{2}(x_{2}-x_{1})^2 + \frac{K_{2}}{2}(x_{2}-x_{3})^2  \nonumber \\ 
& &+\frac{g}{4}(x_{2})^4, \nonumber \\ (V+U)_{3}&=&\frac{K_{0}}{2}(x_{3}-x_{5})^2 + \frac{K_{2}}{2}(x_{3}-x_{2})^2 \nonumber \\ & &+\frac{K_{3}}{2}(x_{3}-x_{4})^2 + \frac{g}{4}(x_{3})^4,  \nonumber \\
(V+U)_{4} &=& \frac{K_{3}}{2}(x_{4}-x_{1})^2 + \frac{K_{3}}{2}(x_{4}-x_{3})^2   \nonumber \\ 
& &+\frac{g}{4}(x_{4})^4.
\end{eqnarray}
Here $K_{2}$ is the spring constant of the two springs connecting $m_2$ to $m_1$ and $m_3$, and $K_{3}$ is the spring constant of the springs connecting $m_4$ to $m_1$ and $m_3$.

The equations of motion when the system is connected to two Langevin reservoirs at temperatures $T_L$ and $T_R$ can be derived accordingly. We will focus on the harmonic case when $g=0$ and tune $m_2,m_4,K_2,K_3$ to look for atypical behavior in the system. For simplicity, we set $m_1/m_3=1 $,  $K_0/K_3=1$, and $b_L=b_R=b$.
Fig.~\ref{fig:foursite} $(b)$ shows the local thermal current flowing through $m_2$ in the steady state when $m_2/m_3=0.4$, $m_4/m_3=0.5$, and $K_2/K_3=0.35$. Apparently, the local atypical thermal current survives in the more complicated setup with no direct coupling between the two masses connected to the Langevin reservoirs. Moreover, the dependence of the local current on the system-reservoir coupling is similar to the result shown in the upper panel of Fig.~\ref{fig:HarmonicPhase}. Therefore, the local atypical thermal current is robust against asymmetry in the system-reservoir coupling, inhomogeneity in the system-substrate coupling, and additions of more masses and springs to the system.

\section{Experimental implications and possible applications}\label{sec:app}
Although the Langevin equation~\eqref{eq:EOM} does not differentiate the size of the system, it may be more feasible to realize the system in molecular or nano-mechanical systems due to the properties of the reservoirs. As shown in Fig.~\ref{fig:HarmonicPhase}, increasing the value of $b$ can switch the direction of the local thermal current. On the other hand, the nonlinear system-substrate coupling has little effect on the direction of the local thermal current, as shown in Fig.~\ref{fig:NonlinearDiagram}.

In principle, all the parameters of the harmonic system and its couplings to the reservoirs and substrate should be tunable. There are studies and techniques for tuning the coupling between a molecular or nano-mechanical system and its environment~\cite{landolsi2013adhesion, nelson2007measuring, ong2011effect, xu2009nanoengineering,si2016strain,ratchford2011manipulating, lee2008measurement,yilbas2001material,li2015mechanical}. 
Tuning techniques are usually classified into two categories: mechanical or electromagnetic. In mechanical methods, one may use the atomic force microscopy \cite{ratchford2011manipulating, lee2008measurement} to locally strain the material. The contacts between the system and the reservoirs can then be modified. For example, using density-functional theory calculations, Ref.~\cite{li2015mechanical} shows that the thermal current in a 
molecular junction can be manipulated through mechanical compression for a wide range of temperatures, essentially due to mode localization. 
Ref.~ \cite{li2017strategy} shows, also using density functional theory, a significant suppression in the phononic thermal conductance of a molecular junction due to its structure. 

On the other hand, electromagnetic methods use tunable, high frequency pulsed laser \cite{yilbas2001material} to adjust the interactions between the system and reservoirs on the nano or atomic scale, thereby changing $b$. Alternatively, for substrates that are conducting or piezoelectric \cite{ding2010stretchable}, an electric current can be sent through the substrate to modify its coupling with the system. The modification will also affect the connections between the system and reservoirs, so the value of $b$ can be tuned indirectly. In this case, it is likely the nonlinear coupling $g$ may be unintentionally modified as well. However, since the system-substrate coupling has minimal influence on the direction of local thermal current as shown in Fig.~\ref{fig:NonlinearDiagram}, tuning properties of the substrate to adjust the system-reservoir coupling $b$ indirectly becomes a viable option.

Since the direction of the local thermal current can be tuned by a variety of techniques, one can envision possible applications of the setup shown in Figs.~\ref{fig:schematic} and \ref{fig:foursite} in molecular or nano-scale mechanical devices. For example, tuning the system-reservoir coupling by mechanical pressing or electromagnetic field as mentioned above can reverse the local thermal current, and one may design a thermal switch in a local region embedded in a multi-path geometry. This will allow an active control of the thermal current through a designated region. 

As another example, one may identify the two directions of the local thermal current (cold-to-hot vs. hot-to-cold) with the two binary digits 0 and 1 and design classical memory elements where the local current is tuned by mechanical or electromagnetic means to perform writing. One may couple additional masses and springs to $m_1$ and $m_2$ to siphon a small amount of the thermal current in order to read out the information. The geometry-based thermal-current switch and memory element complement topological thermal switch and logic~\cite{Pirie18} and enrich the thriving fields of phononics~\cite{li2012colloquium} and heattronics~\cite{chien2013tunable}. 

\section{\label{sec:level4}Conclusion}
In summary, we have demonstrated a steady-state local thermal current from cold to hot in a classical harmonic system coupled to two Langevin reservoirs at different temperatures. The overall thermal current, nevertheless, is always from hot to cold, as required by the second law of thermodynamics. The regime exhibiting the local atypical current is not only determined by the parameters of the harmonic system, but also by the coupling to the reservoirs. Interestingly, there exist parameters where the local thermal current vanishes in the steady state. 
Moreover, the local atypical thermal current is found to be robust against nonlinear effects modeling the system-substrate coupling, asymmetric system-reservoir coupling, and additions of more masses and springs.

While topological effects can lead to interesting quantization of classical thermal transport~\cite{chien2018topological} and thermal logic~\cite{Pirie18}, geometric effects such as the multi-path setup studied here can also give rise to unconventional thermal transport behavior and lead to interesting thermal devices.

\textit{Acknowledgment}: We thank Prof. Linda Hirst, Prof. David Strubbe, and Dr. Kirill A. Velizhanin for stimulating discussions and Dr. Chen-Yen Lai for helping with the simulations. We also acknowledge computing time on the MERCED cluster at UC Merced funded by National Science Foundation Grant No. ACI-1429783.

\bibliographystyle{apsrev}

\end{document}